\documentclass[german,english]{bvm2019}

\title{Dilated deeply supervised networks for hippocampus segmentation in MRI\thanks{Supported by the EFI project: BIG-THERA.}}

\author{Lukas Folle$^1$, Sulaiman Vesal$^1$, Nishant Ravikumar$^1$, Andreas Maier$^1$}

\authorrunning{L. Folle et al.}
\institute{%
$^1$Pattern Recognition Lab, Friedrich-Alexander-Universit\"at Erlangen-N\"urnberg, Germany}

\email{lukas.folle@fau.de}

\begin{document}

%==============================================================================
% w�hlen Sie mit dem Befehl \selectlanguage die Sprache aus, in der Ihr 
% Proceeding verfasst ist
%
\selectlanguage{english}
%\selectlanguage{english}

\maketitle

\begin{abstract}
Tissue loss in the hippocampi has been heavily correlated with the progression of Alzheimer's Disease (AD). The shape and structure of the hippocampus are important factors in terms of early AD diagnosis and prognosis by clinicians. However, manual segmentation of such subcortical structures in MR studies is a challenging and subjective task. In this paper,  we investigate variants of the well known 3D U-Net, a type of convolution neural network (CNN) for semantic segmentation tasks. We propose an alternative form of the 3D U-Net, which uses dilated convolutions and deep supervision to incorporate multi-scale information into the model. The proposed method is evaluated on the task of hippocampus head and body segmentation in an MRI dataset, provided as part of the MICCAI 2018 segmentation decathlon challenge. The experimental results show that our approach outperforms other conventional methods in terms of different segmentation accuracy metrics.  
\end{abstract}

\section{Introduction}
Neurodegenerative brain disorders are a major cause of disability, and early mortality, in many developed and developing countries worldwide. Alzheimer's disease is a type of dementia that affects 20 \% of the population over 80 years of age, worldwide \cite{2718-01}. Currently, AD is typically only diagnosed in patients presenting with symptoms of cognitive impairment, and behavioural changes \cite{2718-03}. With high-resolution MRI structural changes in the brain which accompany the onset of AD, can be recognized in vivo \cite{2718-02}. Early disease stages classified as mild cognitive impairment that occur prior to AD, can also be identified in some cases, and the associated structural changes within the brain can subsequently be used as biomarkers to predict the risk of conversion to AD. Additionally, the rate of tissue atrophy of the hippocampus can be used as a temporal marker to monitor the progression of AD. 
The current clinical protocol to detect volumetric changes in the hippocampus is manual segmentation, which is time-consuming, observer-dependent and challenging \cite{2718-03}. Consequently, an automated approach to hippocampus segmentation is imperative to improve the efficiency and accuracy of the diagnostic workflow. Several automatic and semi-automatic segmentation approaches have been proposed, which utilize T1-weighted structural MRIs, to segment the hippocampus. A multi-atlas segmentation approach was proposed in \cite{2718-10}, to jointly localize and segment the hippocampi using the average of all registered atlases. In \cite{2718-11}, robust segmentation approach was proposed, using subject-specific 3D optimal local maps, with a hybrid active contour model to automatically segment hippocampus.

\begin{figure}
	\centering
	\includegraphics[width=0.90\textwidth]{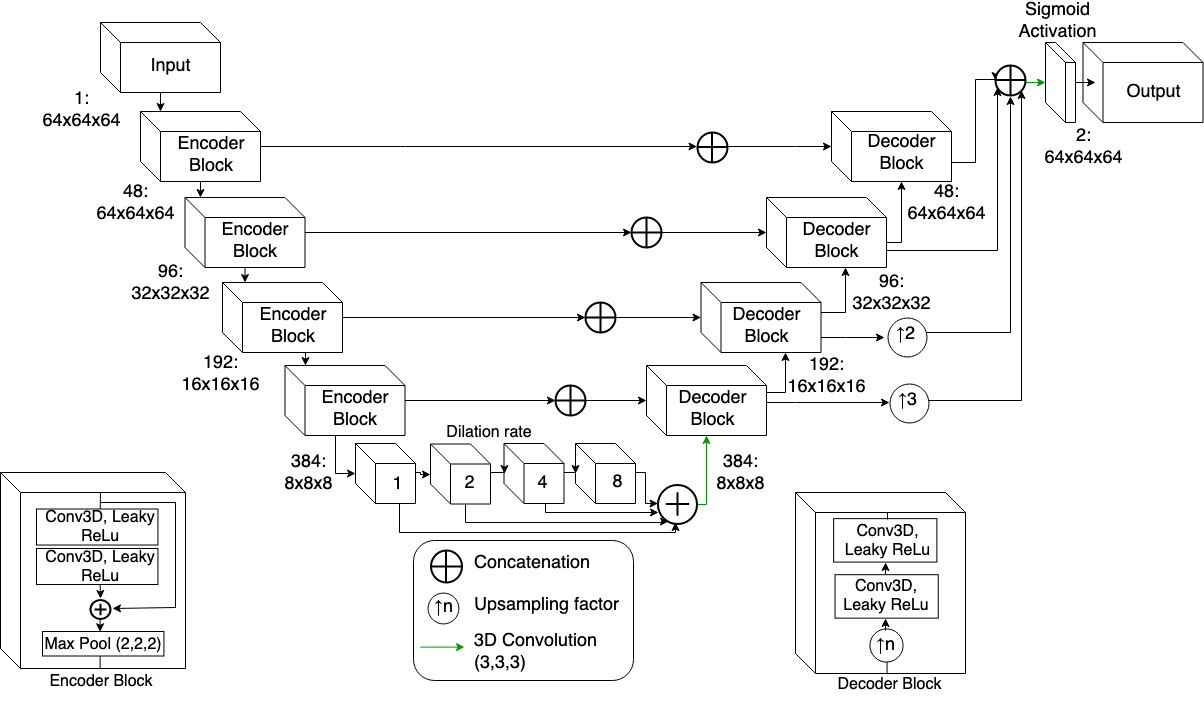}
	\caption{Network Architecture with residual connections in the encoder path, dilated convolutions at the lowest layer and residual connections between the decoder stages and the final layer.}
	\label{fig2}
\end{figure}

In recent years, convolution neural networks (CNNs) have achieved state-of-the-art performance in a variety of medical image segmentation tasks. Specifically, the U-Net \cite{2718-06}, an encoder-decoder network, has received tremendous attention within the medical imaging community. Expanding the U-Net to process 3D volumes rather than 2D slices was proposed in \cite{2718-12} using 3D convolutions (3D U-Net). This was modified in \cite{2718-09} by increasing the channels in the center part of the network (V-Net). In this paper, we propose a CNN for automatic segmentation of the hippocampus. Our network is based on the 3D U-Net, with dilated convolutions in the lowest layer between the encoder and decoder paths \cite{2718-08}, residual connections between the convolution blocks in the encoder path, and residual connections between the convolution blocks and the final layer of the decoder path. A schematic of the network is presented in Fig.~\ref{fig2}. The main contribution of this paper is the combination of dilated convolutions in the lowest part of the network with the ensemble of the decoder outputs for the final prediction, providing a mechanism for `deep supervision'. We evaluated the performance of the network using the hippocampus dataset provided as part of the Medical Segmentation Decathlon challenge\footnote{http://medicaldecathlon.com/} hosted at MICCAI 2018, and compared it to different 3D U-net based architectures.

\section{Methodology}
Segmentation tasks often require integration of local and global context, in addition to learning multi-scale features. However, training segmentation networks that incorporate these properties and act directly on volumetric data, is computationally intensive. We address this by including dilated convolutions within the network to imbue greater global context during feature extraction and combine the output of the decoder layers for the final mask prediction, thereby encouraging the learning of multi-scale features, while providing a means for efficient backpropagation of gradients through the network. Beyond that, this modification yields the benefit of residual connections to the decoding part while retaining the same number of model parameters.

The proposed network consists of four encoder and decoder blocks, each containing two 3D convolution layers with kernel size of 3x3x3, batch normalization and leaky rectified linear units (leaky RELU) as activation functions. The encoder blocks additionally use residual connections and 3D max-pooling operations, as illustrated in Fig.~\ref{fig2}. The decoder blocks use 3D up-sampling with a factor of two. The four dilated convolution layers employed in the bottleneck of the network are configured such that the first layer uses a dilation rate of one, and each subsequent layer increases the dilation rate by a multiple of two, as proposed in \cite{2718-08}. The output of each decoder block is up-sampled to match the dimensions of the final mask predicted by the network, following which, they are all concatenated.
\subsection{Data acquisition}
Images from 263 subjects were provided as part of the Medical Decathlon challenge 2018, for hippocampus head and body segmentation. The subjects were scanned with a T1-weighted MPRAGE sequence (TI / TR / TE = 860 / 8.0 / 3.7ms) and manually annotated with the left and right, anterior and posterior, hippocampus by Vanderbilt University Medical Center. We split the data set such that 90\% were used for training and validating the network, via nine-fold cross-validation, and 10\% of the data-set was used for testing. As the data provided was already truncated to the region of interest around the hippocampus, very little data pre-processing was necessary. Z-score normalization based on mean and standard deviation of the intensities was applied to each patient scan.
\subsection{Training procedures}
Our model is trained from scratch and evaluated using the Dice similarity coefficient (DSC), Jaccard index (JI) and normalized surface distance (NSD). DSC and JI measure the overlap of the ground-truth and model-predicted segmentations, while NSD is computed between the reconstructed surfaces. These were the official metrics used to assess segmentation accuracy in the decathlon challenge as well. The dice coefficient loss is widely used for training segmentation networks \cite{2718-09}. We used a combination of binary cross entropy and DSC loss functions to train all networks investigated in this study, as proposed in \cite{2718-08}. This combined loss (Eq.\ref{eq1c}) is less sensitive to class imbalance and leverages the advantages of both loss functions. Our experiments demonstrated better segmentation accuracy when using the combined loss in comparison to employing either individually.

\begin{equation}
\label{eq1c}
\zeta(y, \hat{y}) = \zeta_{dc}(y, \hat{y}) + \zeta_{bce}(y, \hat{y})
\end{equation}

In Eq.(\ref{eq1c}) $\hat{y}$ denotes the output of the model and the ground truth labels are denoted by $y$. We use the two-class version of the DSC loss $\zeta_{dc}(y, \hat{y})$ proposed in \cite{2718-09}\cite{2718-08}, the Adam optimizer with a learning rate of 0.0005, and trained the network for 500 epochs. Additionally, the learning-rate was reduced gradually (using a factor of 0.1), if the validation loss did not improve after 10 epochs. To prevent overfitting and improve the robustness of our approach to varied hippocampal shapes, we augmented the dataset with random rotations and flipping. Based on our experiments, we found that augmenting with large rotation angles produced worse segmentation masks, consequently, we reduced the rotation angles to be in the range of $\pm 10$ degrees. 
%All networks investigated in this study were trained using the same hyper-parameters.

\section{Results and Discussion}
In order to assess the performance of different networks, we used the Dice Coefficient Score (DSC), Jaccard index and Normalised Surface Distance (NSD) with 4mm tolerance. The segmentation performance of our model, V-Net, 3D U-Net and 3D U-Net with dilated convolutions are compared in Table~\ref{tab1}. The V-Net achieved mean DSC scores of 96.8\%, 87.2\% and 84.8\%, for the training, validation and test sets, respectively. The performance of the 3D U-Net is close to the V-Net performance with mean DSC scores of 96.5\%, 85.5\% and 86.5\% respectively. 3D U-Net with dilated convolutions was able to improve the scores to 97.7\%, 87.8\% and 87.9\%, respectively. However, the proposed approach outperformed the others with scores of 98.4\%, 88.2\% and 88.2\% for the training, validation, and test sets, respectively. Additionally, our approach consistently outperformed the other state-of-the-art networks, in terms of the JI and NSD metrics as well, as highlighted in Table~\ref{tab1}.

\begin{figure}[h]
	\centering
	\includegraphics[width=0.325\textwidth]{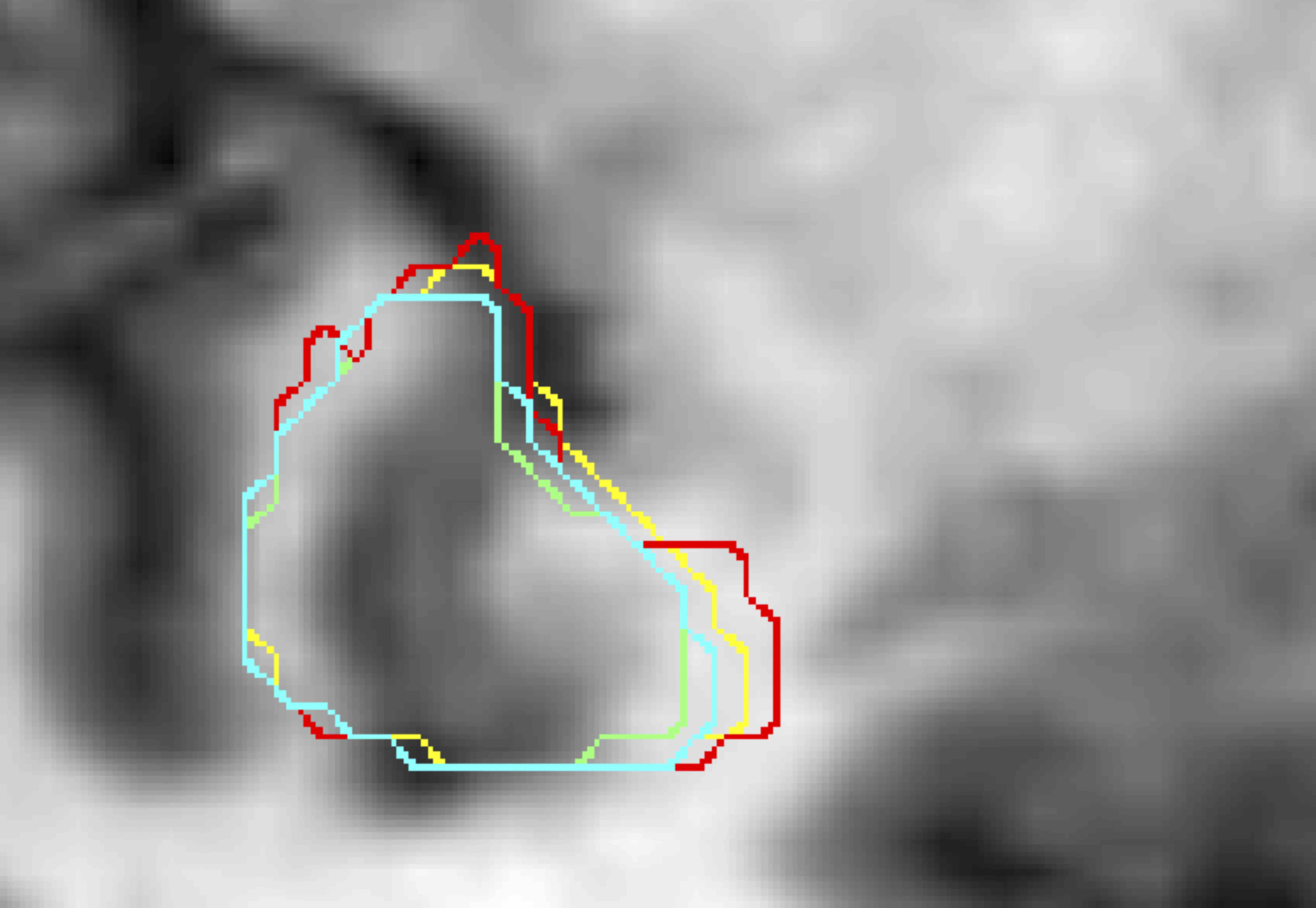}
	\includegraphics[width=0.325\textwidth]{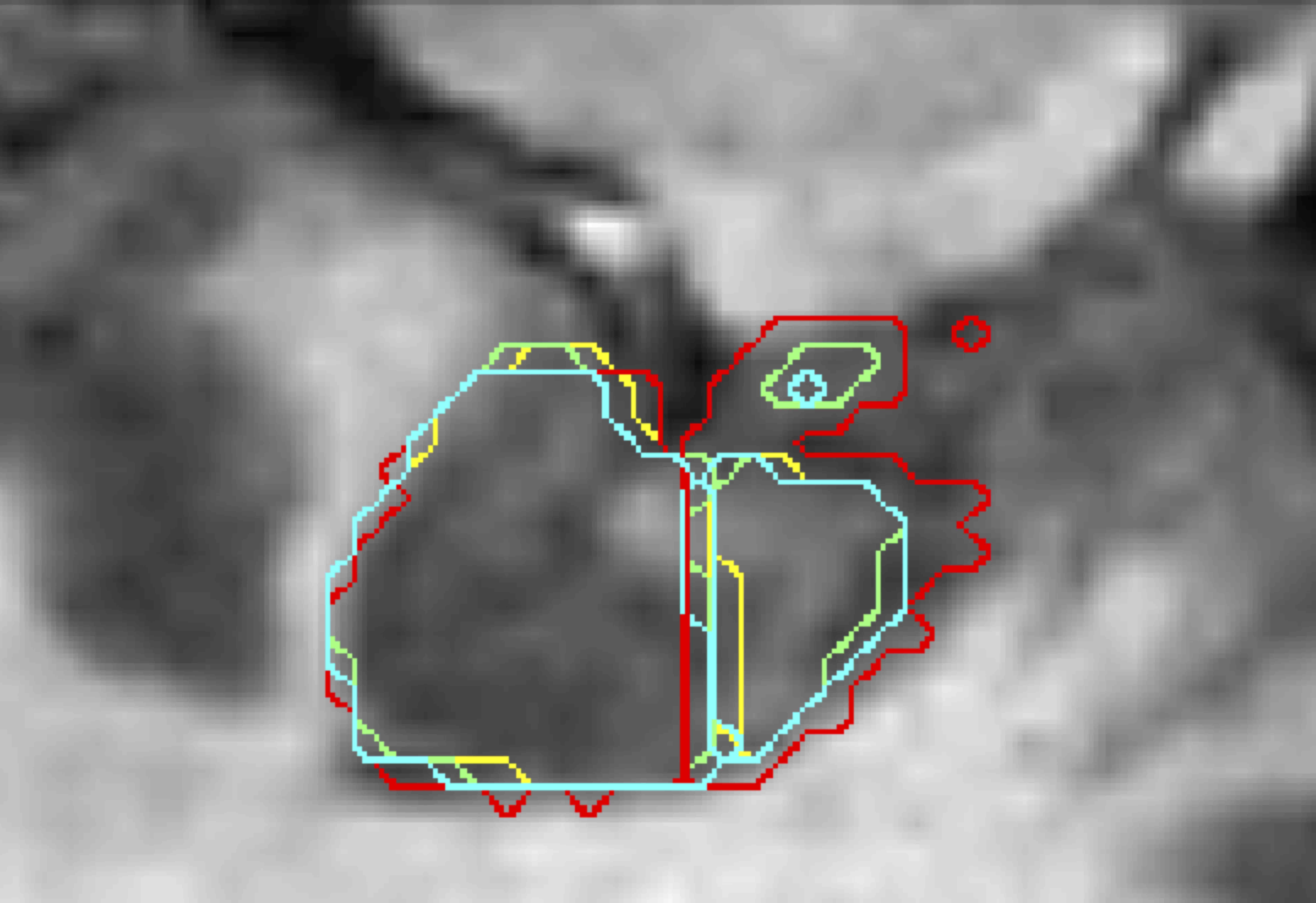}
	\includegraphics[width=0.325\textwidth]{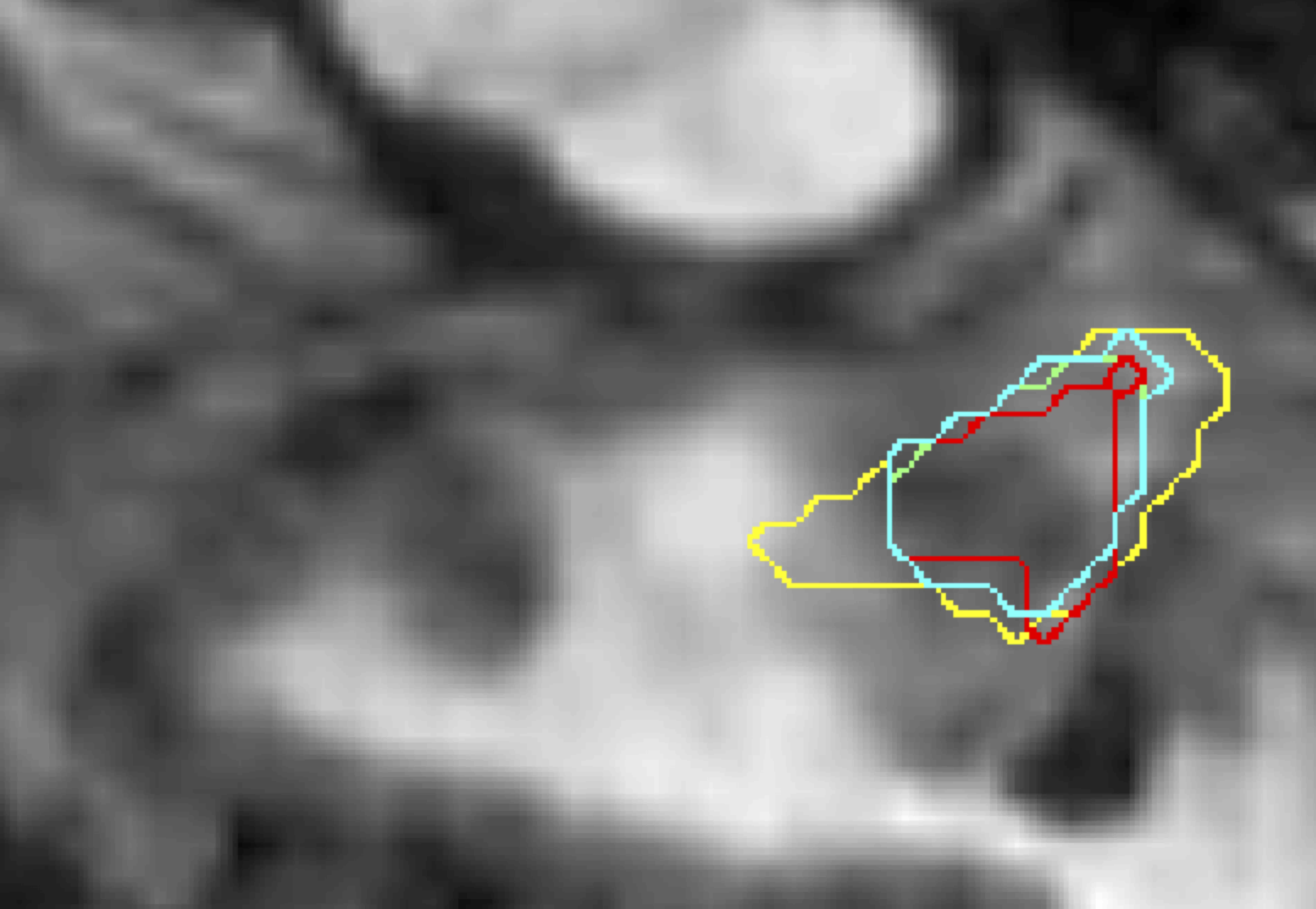}
	\caption{Each image represents a different MRI slice from a different patient. The corresponding segmentations are overlaid: Red contour represents ground-truth, yellow V-Net, green 3D U-Net with dilated convolutions and cyan our proposed method.}
	\label{fig:slices}
\end{figure}

\begin{table}[htbp]%htbp
	\centering
	\caption{Segmentation accuracy evaluated in terms of DSC, JI and NSD for the V-Net, 3D U-Net, dilated 3D U-Net and the proposed method.}
	\begin{tabular}{l|lllll}
		 & Training  &  Validation & \multicolumn{3}{c}{Testing} \\\cline{1-6}
		Methods & DSC   & DSC   & DSC   & JI    & NSD \\\cline{2-6}
		V-Net & 0.968 & 0.872 & 0.848 & 0.736 & 0.954 \\
		3D U-Net & 0.965 & 0.858 & 0.865 & 0.740 & 0.960 \\
		3D U-Net + Dilation & 0.977 & 0.878 & 0.879 & 0.785 & 0.960 \\
		Proposed method & 0.984 & 0.882 & 0.882 & 0.790  & 0.962 \\
	\end{tabular}%
	\label{tab1}%
\end{table}%

\begin{figure}[htb]
	\setlength{\figbreite}{0.3\textwidth}
	\centering
	\caption{Rows represent 3D surface visualizations for two different patients. Columns from left to right are the ground truth surfaces, and those predicted by the V-Net and our approach, respectively. Hippocampus head is visualized in green and body in red.}
	\label{fig:volumes}
	\rotatebox{90}{Test case 1}
	\subfigure[Ground truth]{\includegraphics[width=\figbreite]{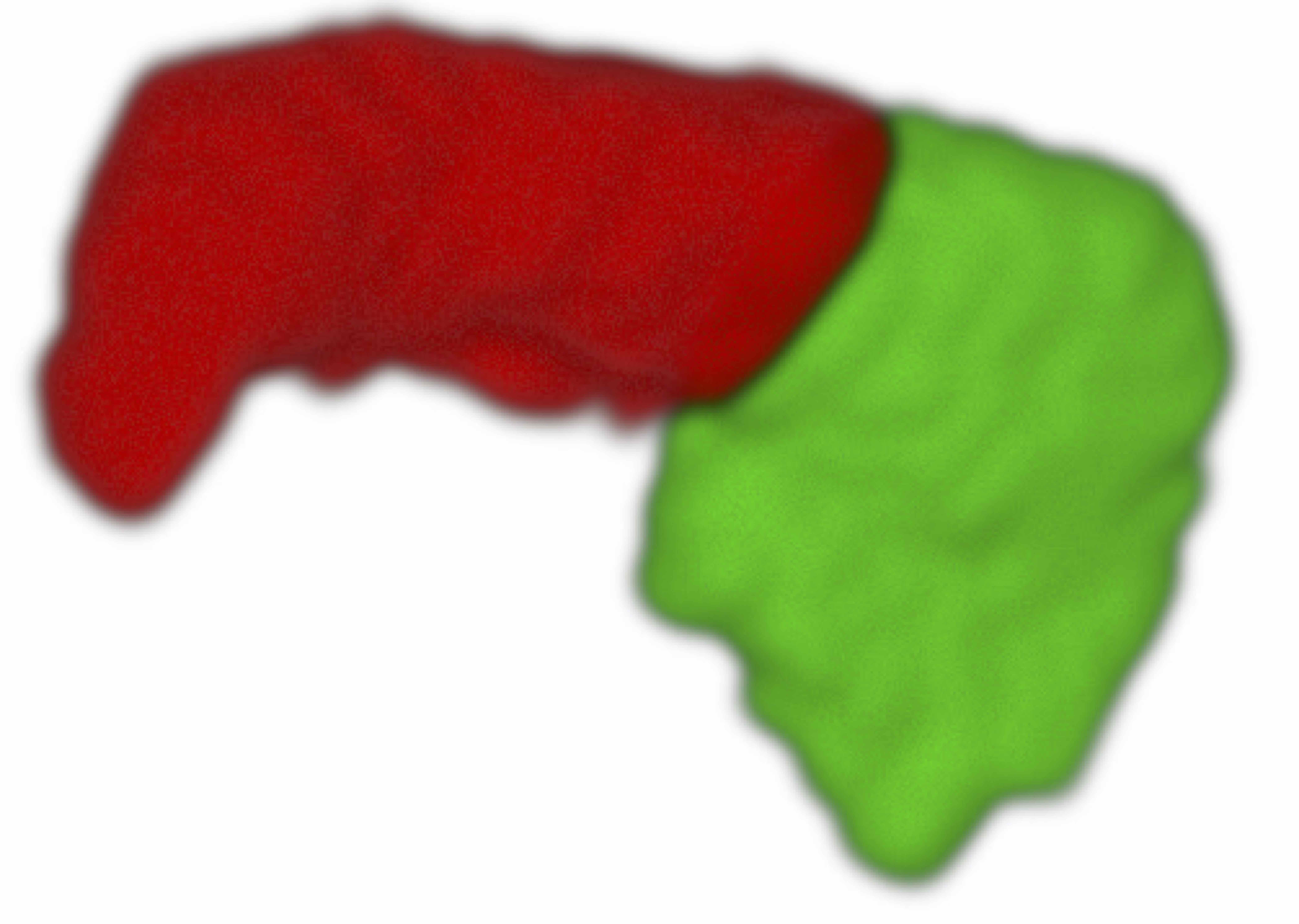}}
	\subfigure[V-Net]{\includegraphics[width=\figbreite]{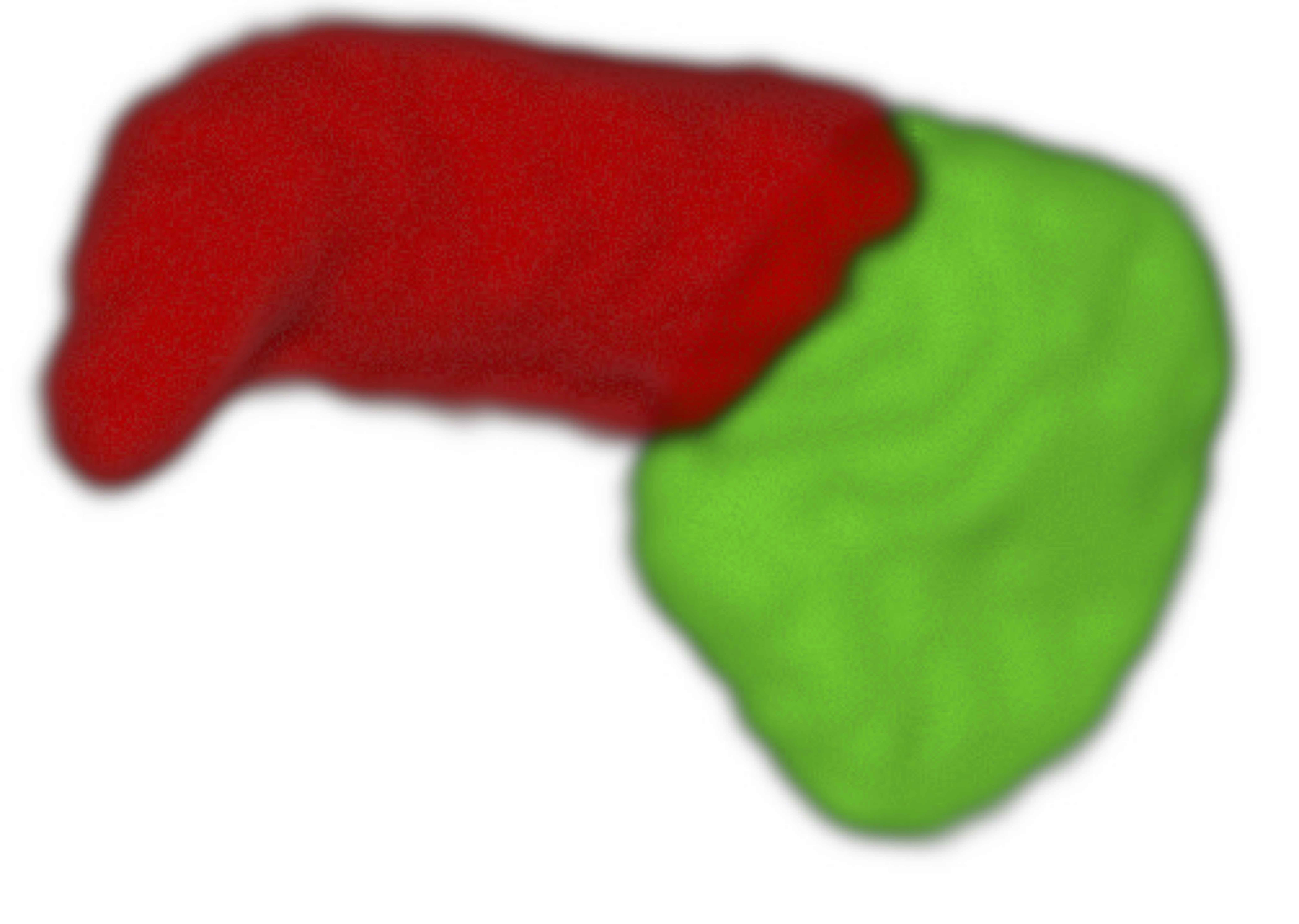}}
	\subfigure[Proposed method]{\includegraphics[width=\figbreite]{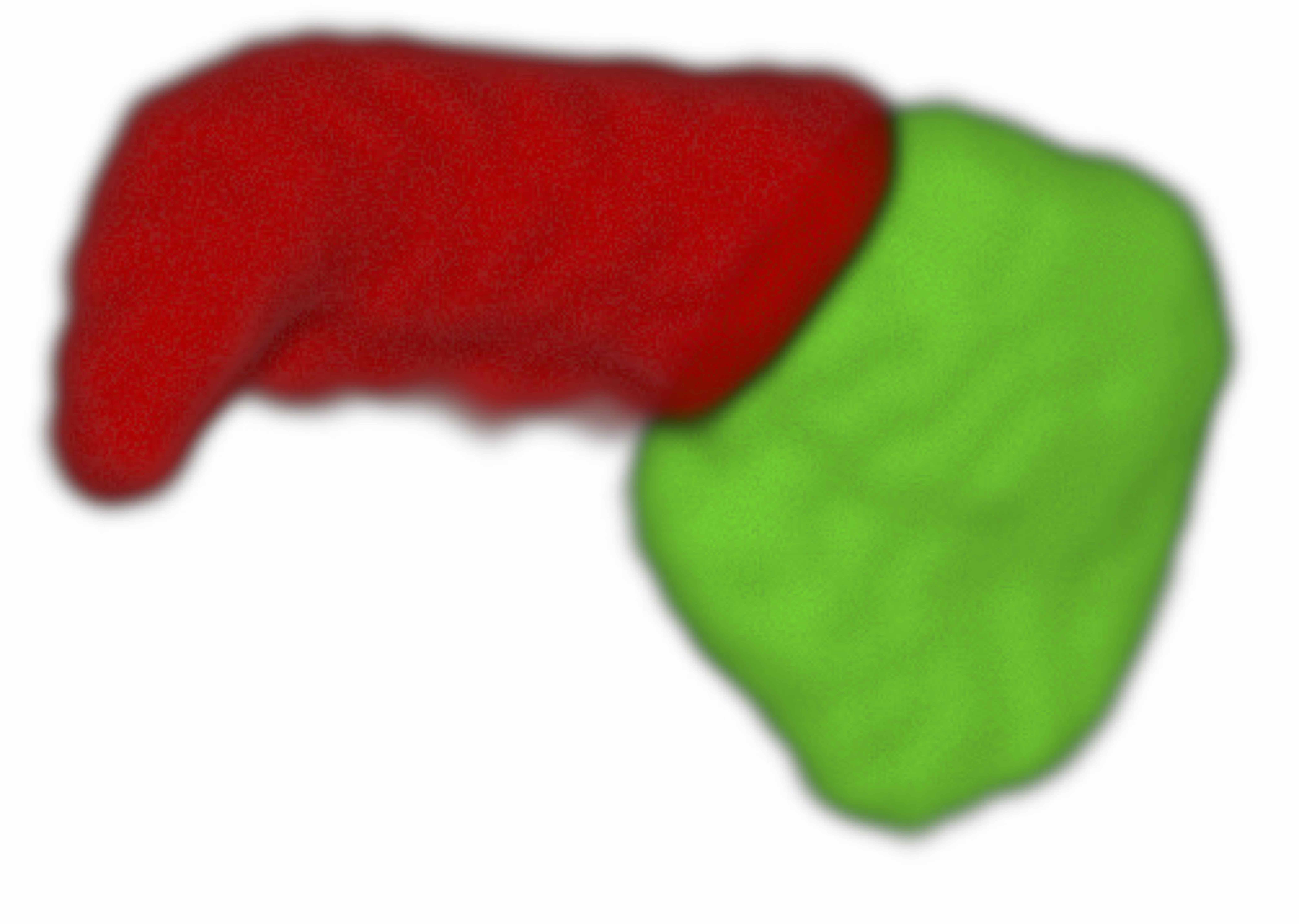}}
	\par\medskip
	\rotatebox{90}{Test case 2}    
	\subfigure[Ground truth]{\includegraphics[width=\figbreite]{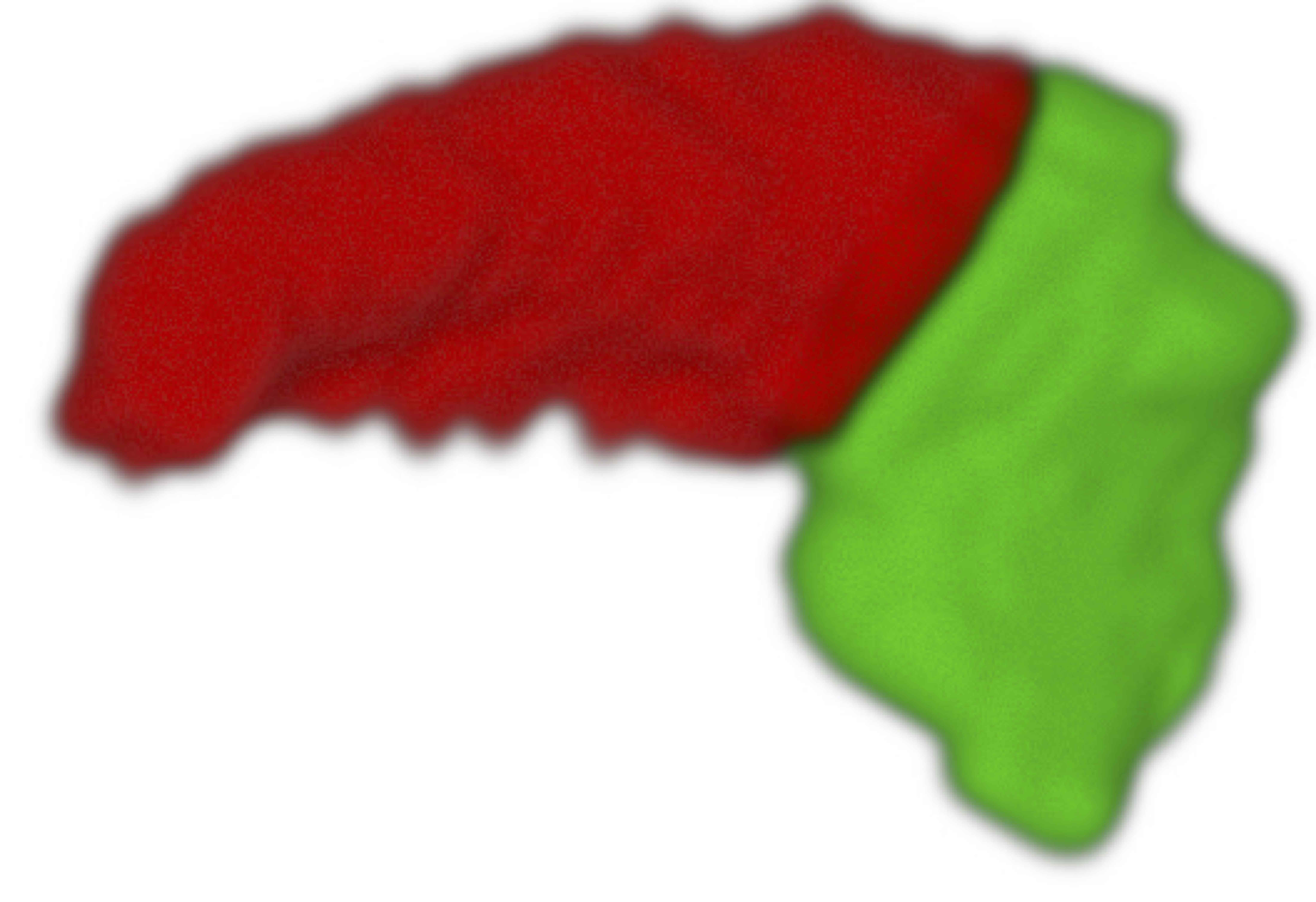}}
	\subfigure[V-Net]{\includegraphics[width=\figbreite]{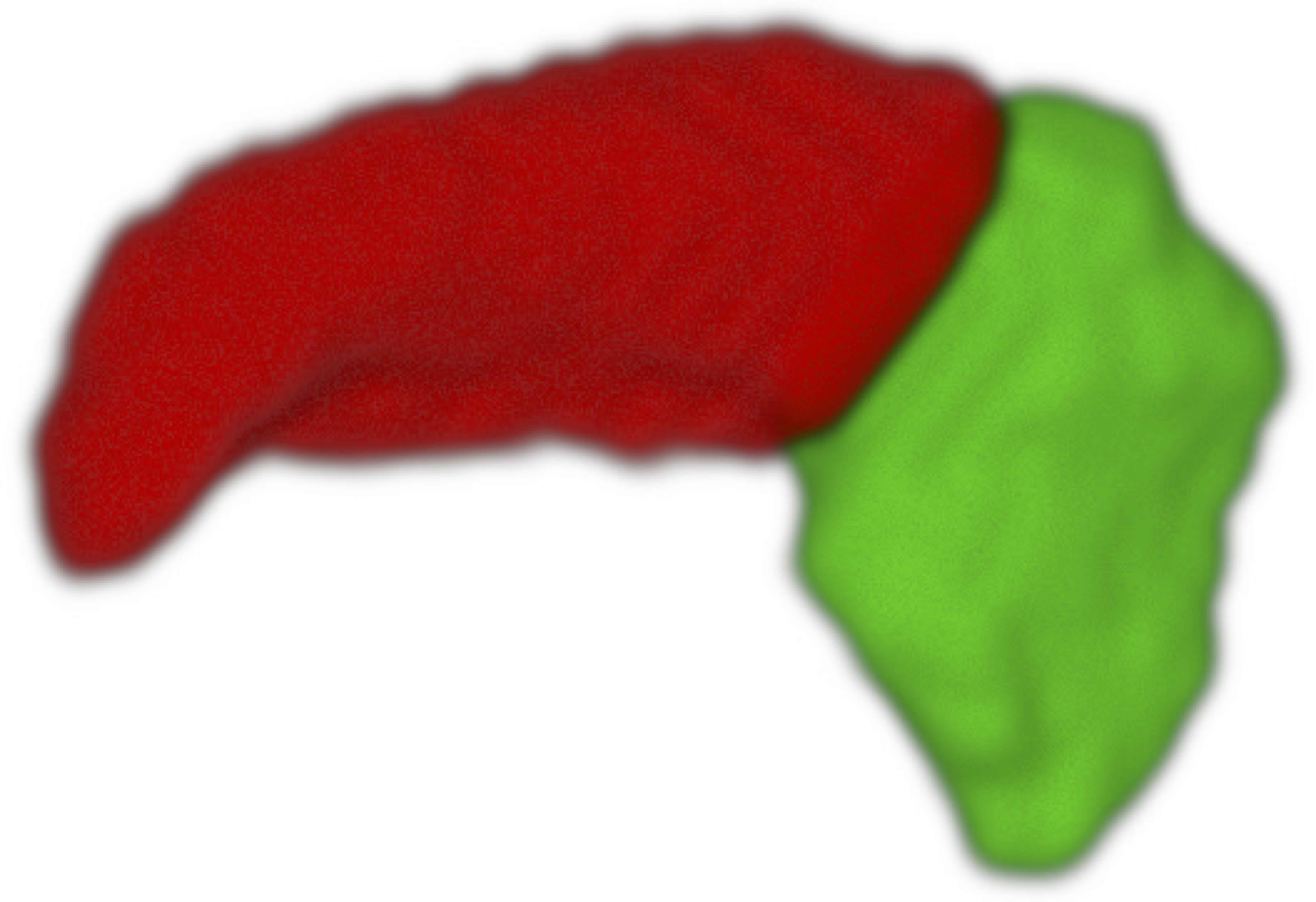}}
	\subfigure[Proposed method]{\includegraphics[width=\figbreite]{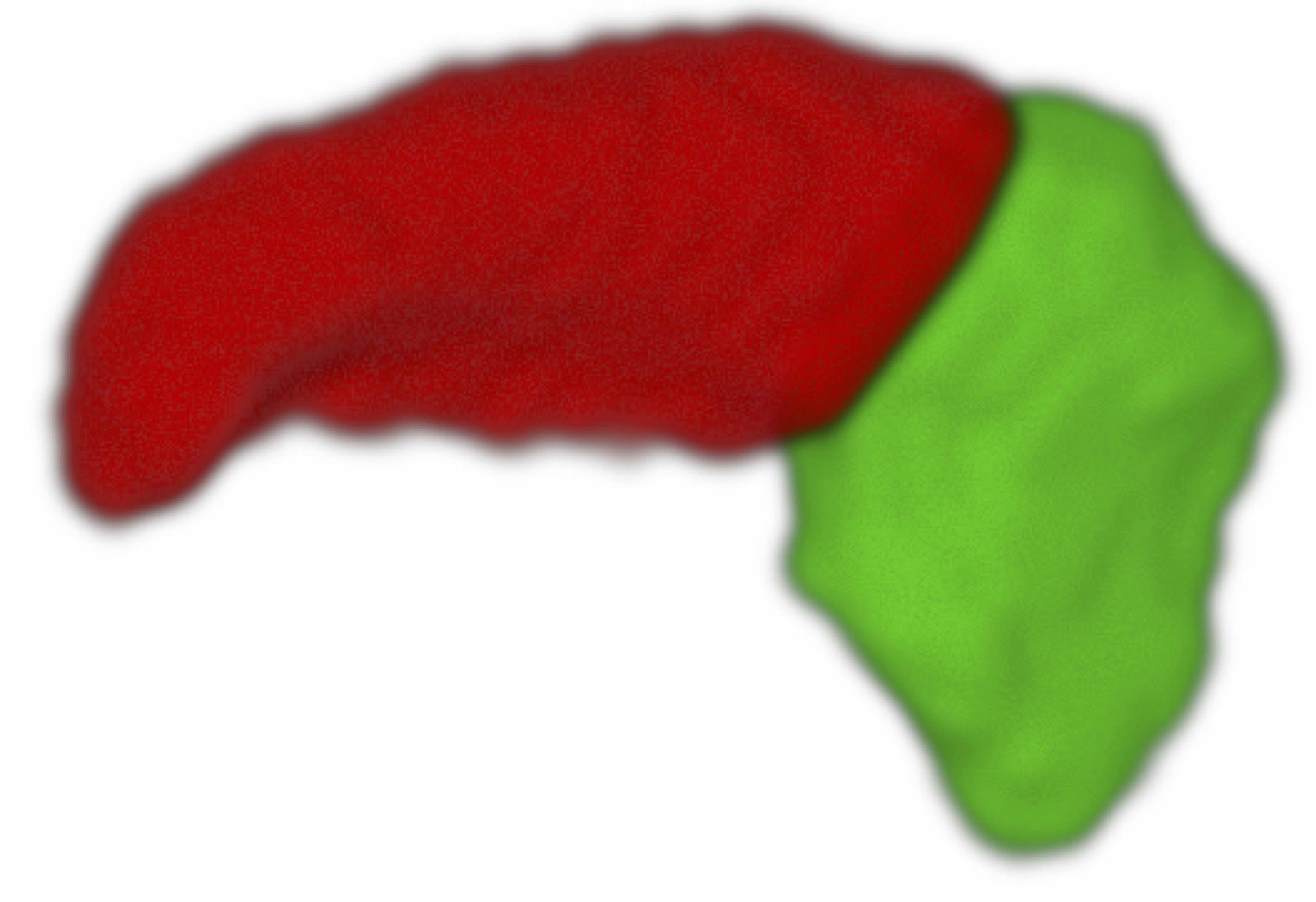}}

	\par\medskip
\end{figure}

In Fig.~\ref{fig:slices} the segmentation quality of the proposed method is visually compared with V-Net, 3D U-Net, and 3D U-Net with dilated convolutions. Here, red represents the ground-truth, yellow, green and cyan represent the predictions of V-Net, dilated 3D U-Net and the proposed method, respectively. In the second column, the advantage of dilated convolutions is highlighted, in comparison to the V-Net, which failed to segment the small disjoint parts of the mask in the top right. However, the dilated 3D U-Net and the proposed method were able to capture those areas due to the increased global context imbued in the learned features. Fig.~\ref{fig:volumes} depicts 3D surface meshes of two different patients. Columns two and three illustrate the outputs of V-Net and our method, respectively. The lower boundary of the red part (body) of the hippocampus in the ground-truth surfaces, contains ridge-like structures which are typical of hippocampal structure. While the V-Net predicted surfaces are relatively smooth in this region, the proposed approach is more successful in capturing these subtle shape variations.

\section{Conclusion}
We proposed a 3D U-Net based segmentation framework with dilated convolutions in the deepest part of the network and deep supervision in the decoder part of the network. The dilated convolutions capture global context due to their larger receptive fields. Deep supervision helped further improve segmentation accuracy, by incorporating multi-scale information more efficiently during the training process. We showed that our network consistently outperforms the V-Net, 3D U-Net, and 3D U-Net with dilated convolutions, in terms of all metrics evaluated. Future work will aim to use the proposed framework for segmentation in whole brain MRI volumes, and on different segmentation tasks in medical imaging.

\bibliographystyle{bvm2019}

\bibliography{2718}
% Bitte setzen Sie hier Ihre Beitragsnummer ein und benennen Sie
% die BibTeX-Datei ebenfalls auf Ihre Beitragsnummer um.
%Kontrollzeiledef
\end{document}